\documentclass[conference,a4paper]{IEEEtran}
\IEEEoverridecommandlockouts

\usepackage{cite}
\usepackage{amsmath,amssymb,amsfonts}
\usepackage{graphicx}
\usepackage{booktabs}
\usepackage{xcolor}
\usepackage{url}
\urldef{\fastmlurl}\url{https://fastmachinelearning.org}
\usepackage{tikz}
\usepackage{xspace}
\usepackage{enumitem}
\usetikzlibrary{positioning,calc}
\tikzset{
  io/.style    = {draw, rounded corners=2pt, inner sep=3pt, font=\scriptsize, fill=black!5},
  dense/.style = {draw, inner sep=3pt, align=center, font=\scriptsize, fill=olive!20},
  dw/.style    = {draw, inner sep=3pt, align=center, font=\scriptsize, fill=red!12},
  act/.style   = {draw, inner sep=3pt, font=\scriptsize, fill=yellow!25},
  lay/.style   = {draw, inner sep=4pt, font=\scriptsize, fill=red!12},
  dots/.style  = {font=\scriptsize},
  ar/.style    = {-latex, semithick},
}
\usetikzlibrary{arrows.meta,positioning,calc,decorations.pathreplacing}

\graphicspath{{figs/}}

\newcommand{\pwconv}{PW\raisebox{1pt}{$\circledast$}\xspace}
\newcommand{\dwconv}{DDW\raisebox{1pt}{$\circledast$}}

\setlist[description]{
    font=\bfseries,
    leftmargin=0pt,
    parsep=\parsep,
    listparindent=\parindent,
    labelwidth=0em,
    itemindent=1em,
    labelsep=1em,
    align=left,
    itemsep=\parsep,
}

\begin{document}
\bstctlcite{IEEEexample:BSTcontrol}

\title{Realizing Low-Precision Storage and Sparse Execution in a Folded FPGA Speech Enhancer\thanks{This work was funded by the European Union through the Chips Joint Undertaking project NeAIxt (grant agreement No.~101194172). The author thanks the FastML community (\fastmlurl) for the open-source hls4ml and HGQ tools and for feedback on the manuscript.}}

\author{\IEEEauthorblockN{Clément Laroche}
\IEEEauthorblockA{Audio Research, GN A/S, Ballerup, Denmark\\
Email: claroche@gn.com}
}

\AddToHookNext{shipout/foreground}{\put(0.5\paperwidth,-0.9cm){\makebox[0pt][c]{\parbox[t]{0.8\paperwidth}{\centering\footnotesize This work has been submitted to the IEEE for possible publication. Copyright may be transferred without notice, after which this version may no longer be accessible.}}}}
\maketitle

\begin{abstract}
Hardware-aware training often optimizes FPGA resources, latency, or bit operations, but folding changes what compression saves: shared arithmetic decouples precision, weight storage, and scheduled work. We derive training objectives from the measured costs of a folded streaming speech-enhancement accelerator. Mixed-precision weights reduce on-chip storage; structured channel sparsity removes entire operations from the engine schedule. We deploy a bit-exact implementation on the Kria KV260. The best sparse model eliminates \(47.3\%\) of convolution MACs, cutting frame time by \(45\)\% (\(1.86\) to \(1.03\) ms) and dynamic energy by \(60\)\% (\(83\) to \(33\) µJ/frame) against a storage-priced model measured in the same session, with higher speech quality. Relative to the most efficient dense build measured separately, frame time falls by \(42\)\% and energy by \(48\)\%. Reducing weight width from 8 to 6 bits lowers dynamic power by \(27\)\% but changes frame time by less than \(5\)\%. Plain channel group-lasso removes more work than cycle-cost-weighted lasso. %
\end{abstract}

\begin{IEEEkeywords}
speech enhancement, FPGA, mixed precision, structured sparsity, hardware--software co-design, edge inference
\end{IEEEkeywords}

\section{Introduction}\label{sec:intro}

Speech enhancement runs in hearing aids, audio wearables, and voice-activated systems. Responsiveness, privacy, and offline operation put these models on the device rather than in the cloud, which has driven a line of lightweight architectures~\cite{rnnoise,deepfilternet,lisennet}. Beyond architectural design, quantization offers another way to reduce the memory and computational costs of on-device inference~\cite{jacob2018quantization}. Recent work makes quantization itself trainable: mixed-precision searches assign a width per layer under a hardware cost~\cite{haq}, differentiable schemes learn widths and scales jointly with the task~\cite{fracbits,diffq}, and the finest schemes learn a separate width for every channel and every weight~\cite{autoqkeras,hgq}. Commercial silicon, however, prices none of that granularity. A fixed-datapath int8 NPU~\cite{vitisai,stm32n6} pushes every operand through the same $8\times8$ multiplier array. %
Collecting the gains of sub-byte quantization therefore takes hardware built for the trained widths themselves.

Dataflow compilers build exactly that datapath: FINN~\cite{finn,finnr} and hls4ml~\cite{hls4ml,Schulte:2025mai,Aarrestad:2021zos,Ghielmetti:2022ndm} map quantized layers to dedicated hardware, realizing low precision directly in the synthesized arithmetic, while research accelerators instead support multiple arithmetic precisions within the accelerator~\cite{stripes,bitfusion,filmqnn}. These approaches suit latency-critical deployments where much of the network is spatially realized in fabric. For streaming speech enhancement on a small device, the frame budget is generous, but the area is not, so one datapath must serve the whole network in turn. %
Folding changes how compression maps to hardware cost. Here, folding means reusing one physical compute engine for the network's blocks and processing them in sequence instead of giving each block its own hardware. Bit-operation objectives such as Bayesian Bits~\cite{vanbaalen2020bayesianbitsunifyingquantization}, HGQ~\cite{hgq}, and PQuantML~\cite{niemi2026pquantmltoolendtoendhardwareaware} capture implementations in which reducing precision reduces arithmetic work or instantiated logic. Prior single-engine accelerators already show that this correspondence depends on the datapath: layer computations may be time-shared on one processing engine~\cite{dnnweaver,angeleye}, and mixed-precision overlays can retain a fixed-width multiplier while changing only the numerical representation~\cite{que2021circuit}. Thus, no single bit-operation count determines the cost of a folded accelerator.

In this paper, we use the HGQ--HLS4ML flow~\cite{hgq,fastml_hls4ml} to study how learned precision and sparsity translate to a folded FPGA accelerator for streaming speech enhancement. One fixed-width engine is reused across the network~\cite{angeleye,dnnweaver,que2021circuit}; learned widths determine how weights are packed, while exact-zero channel rows are removed from the MAC schedule. A compact table-and-tape representation carries these choices through standard FPGA tools, while bit-exact checks ensure that the deployed accelerator reproduces the trained arithmetic, closing the hardware gap between trained and deployed models~\cite{barbe2026hatorch}. We then adapt the training objective to weight storage and scheduled cycles, following the broader principle that compression should target hardware-specific costs rather than proxy operation counts~\cite{energyprune,netadapt,halp}. Every build exceeds the quality of the same-board 16-bit reference~\cite{olalere2026}, while using $20$--$21$ rather than $899$ DSPs.

\section{The speech-enhancement network}\label{sec:context}\label{ssec:network}

\begin{figure}[t]
\centering
\resizebox{0.88\linewidth}{!}{%
\begin{tikzpicture}[node distance=0.55em]
  \node [io]                       (in)  {$|X_l|$};
  \node [dense, right=of in]       (fe)  {pointwise\\$257{\to}128$};
  \node [lay,   right=1.4em of fe](b1)  {$\mathcal{B}_1$};
  \node [dots,  right=0.5em of b1] (bd)  {$\cdots$};
  \node [lay,   right=0.5em of bd] (bn)  {$\mathcal{B}_{9}$};
  \node [dense, right=1.4em of bn] (be)  {pointwise\\$128{\to}257$};
  \node [act,   right=3.4em of be] (sg)  {$\sigma$};
  \node [io,    right=of sg]       (out) {$\widehat{M}_l$};
  \foreach \a/\b in {in/fe,fe/b1,b1/bd,bd/bn,bn/be,sg/out}
    \draw [ar] (\a) -- (\b);
  \draw [ar] (be) -- node [above, font=\tiny, pos=0.34] {$257$ logits} (sg);
  \coordinate (cut) at ($(be)!0.80!(sg)$);
  \draw [densely dashed, thick, gray] ($(cut)+(0,2.4em)$) -- ($(cut)-(0,2.4em)$);
  \node [font=\scriptsize\itshape] at ($(cut)+(0,3.1em)$) {PL $\mid$ PS};
  \node [dense, below=3.0em of b1, xshift=-2.6em]  (p1) {pointwise\\$128{\to}256$};
  \node [dw,    right=0.5em of p1]  (dwc){depthwise dilated\\$d = 1, 2, 4$};
  \node [dense, right=0.5em of dwc] (p2) {pointwise\\$256{\to}128$};
  \node [draw, circle, inner sep=1pt, right=1.4em of p2, font=\scriptsize] (add) {$+$};
  \draw [ar] (p1) -- (dwc);  \draw [ar] (dwc) -- (p2);  \draw [ar] (p2) -- (add);
  \coordinate (bin) at ($(p1.west) - (1.4em,0)$);
  \coordinate (skip) at ($(p1.south) - (0,1.6em)$);
  \draw [ar] (bin) -- (p1.west);
  \draw [semithick] (bin) |- (skip);
  \draw [ar] (skip) -| (add.south);
  \draw [semithick, densely dashed, gray] (b1.south west) -- (p1.north west);
  \draw [semithick, densely dashed, gray] (bn.south east) -- (p2.north east);
  \node [left=0.5em of bin, font=\scriptsize] {$\mathcal{B}_i$:};
\end{tikzpicture}}
\caption{Conv-FSENet Network. The accelerator returns $257$ logits, and the host applies $\sigma$.}
\label{fig:arch}
\end{figure}
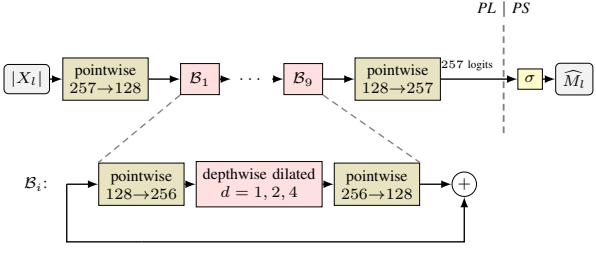

Conv-FSENet~\cite{miccini2025scalable,miccini2025diet} is a Conv-TasNet-style~\cite{convtasnet} dilated temporal convolutional network (TCN) that estimates a mask from STFT magnitude frames (Fig.~\ref{fig:arch}). A pointwise frontend with ReLU feeds nine TCN blocks, arranged as three stacks of three, and a pointwise mask layer. Each block expands pointwise, applies a dilated depthwise \texttt{Conv1D} (kernel $3$; dilation $1$, $2$, $4$ within a stack), projects back pointwise, and adds the block input; BatchNorm and ReLU follow each of the first two convolutions. All convolutions are causal (left-padded) with zero lookahead. The network holds $\sim\!0.66$\,M weights, costs $\sim\!668$\,k MACs per frame, and streams frame by frame. For $L$ frames, $C_\text{res}{=}128$ residual and $C_\text{conv}{=}256$ expanded channels, and kernel size $k{=}3$, a block's pointwise (\pwconv) and dilated depthwise (\dwconv) convolutions cost approximately
\begin{equation}\label{macestimate}
\text{MAC}_\text{\,\pwconv} \approx L \cdot C_\text{conv} \cdot C_\text{res}
\qquad
\text{MAC}_\text{\,\dwconv} \approx L \cdot C_\text{conv} \cdot k .
\end{equation}

\section{The folded engine}\label{sec:system}

\begin{figure}[t]
\centering
\resizebox{0.94\linewidth}{!}{%
\begin{tikzpicture}[
  font=\scriptsize,
  blk/.style = {draw, rounded corners=1pt, minimum height=1.3em, minimum width=1.9em,
                fill=red!12, inner sep=1pt},
  eng/.style = {draw, very thick, rounded corners=2pt, fill=red!12, inner sep=3pt,
                align=center},
  mem/.style = {draw, rounded corners=1pt, fill=olive!20, inner sep=2pt, align=center},
  tbl/.style = {draw, rounded corners=1pt, fill=yellow!25, inner sep=2pt, align=center},
  st/.style  = {draw, rounded corners=1pt, fill=black!5, inner sep=2pt, align=center},
  io/.style  = {draw, rounded corners=2pt, fill=black!5, inner sep=2pt},
  ar/.style  = {-latex, semithick},
]

\node[io] (ain) at (0,0) {$|X_l|$};
\node[blk, right=0.5em of ain] (afe) {fe};
\node[blk, right=0.5em of afe] (a1)  {$\mathcal{B}_1$};
\node[blk, right=0.5em of a1]  (a2)  {$\mathcal{B}_2$};
\node[right=0.5em of a2]       (ad)  {$\cdots$};
\node[blk, right=0.5em of ad]  (a9)  {$\mathcal{B}_9$};
\node[blk, right=0.5em of a9]  (amk) {mask};
\node[io, right=0.5em of amk]  (aout){$\widehat{M}_l$};
\foreach \a/\b in {ain/afe,afe/a1,a1/a2,a2/ad,ad/a9,a9/amk,amk/aout}
  \draw[ar] (\a)--(\b);
\draw[decorate,decoration={brace,amplitude=2.5pt,mirror},thin]
  ([yshift=-2pt]a1.south west) -- ([yshift=-2pt]a9.south east);
\coordinate (mid) at ($(a1.south)!0.5!(a9.south)$);

\draw[ar, very thick] ($(mid)+(0,-0.36)$) -- ($(mid)+(0,-0.84)$);
\node[font=\scriptsize\itshape, anchor=west] at ($(mid)+(0.12,-0.60)$) {fold};
\node[font=\scriptsize, anchor=east] at ($(mid)+(-0.12,-0.60)$)
  {9 iterations; residual copied forward};

\node[eng, text width=17.6em, anchor=north] (engine) at ($(mid)+(0,-1.33)$)
  {\textbf{one block engine} --- $K$ MAC lanes, one shared requantizer\\[1.5pt]
   conv$1{\times}1\to$ BN $\to$ ReLU $\to$ dconv $\to$\\ BN $\to$ ReLU
   $\to$ conv$1{\times}1\to{+}$\\[1.5pt]
   \emph{each tensor: one II$=1$ loop over its \textbf{active} output rows}\\
   \emph{a $b_f{=}0$ row: no tape bits, one tick instead of $128$}};

\node[blk, left=0.8em of engine]  (bfe)   {fe};
\node[blk, right=0.8em of engine] (bmask) {mask};
\draw[ar] (bfe) -- (engine);
\draw[ar] (engine) -- (bmask);
\node[anchor=south, font=\scriptsize\itshape, align=center] at ([yshift=0.25em]bfe.north)   {own\\lanes};
\node[anchor=south, font=\scriptsize\itshape, align=center] at ([yshift=0.25em]bmask.north) {own\\lanes};

\draw[ar, very thick]
  ([xshift=-2.0em]engine.north east) .. controls +(0,0.65) and +(0,0.65) ..
  ([xshift=2.0em]engine.north west);

\node[st, anchor=north, text width=5.0em] (state)
  at ([yshift=-0.95em]engine.south)
  {\textbf{state}\\
   dconv ring,\\
   one bank\\ per block,\\
   one shared head};
\node[mem, anchor=north east, text width=8.6em] (tape)
  at ([xshift=-0.5em]state.north west)
  {\textbf{weight lane}\\
   packed tape, $b_f$ bits/weight\\
   body in BRAM,\\ small in LUT-ROM\\
   width $+$ shift table per row\\
   $K$ codes per cycle};
\node[tbl, anchor=north west, text width=7.0em] (cast)
  at ([xshift=0.5em]state.north east)
  {\textbf{requantization}\\
   one unit, reused;\\ $75$-row cast ROM\\ gives output width,\\
   rounding, overflow\\ per op};

\draw[ar] (tape.north)  -- (tape.north |- engine.south);
\draw[ar, densely dashed] (state.north) -- (engine.south);
\draw[ar] (cast.north)  -- (cast.north |- engine.south);

\end{tikzpicture}}
\caption{Conv-FSENet folded onto one time-shared engine. Block-specific weights arrive on the weight lane, block-specific history from the dconv ring, and the trained output format of every operation from the cast ROM.} %
\label{fig:fold}
\end{figure}
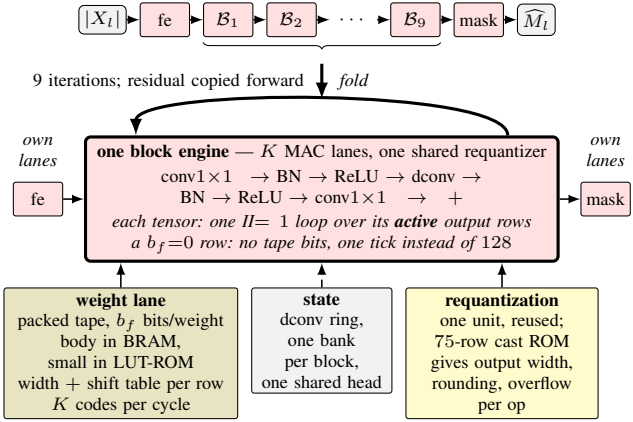

\subsection{Trading latency margin for hardware reuse}

The $16$\,ms frame period provides substantial time for hardware reuse. Our first attempt to implement Conv-FSENet using the HGQ$\to$hls4ml spatial flow exceeded the FPGA's resources. Even with only $1/8$ of the production channel width, the generated design required $356$\,k LUTs, about $3\times$ the device's $117$\,k. hls4ml can reuse hardware within a layer, but it generates separate hardware for each layer, so the resource costs add up across the network. We instead exploit the available cycle budget by time-sharing one physical engine across the network~\cite{finnr}. This replaces the area-heavy spatial mapping with a folded design whose main costs are the shared datapath, the stored weights, and the cycles needed to execute the layer schedule.

This reuse is possible because Conv-FSENet's body contains nine TCN blocks with the same structure; only their weights and depthwise-convolution dilation differ (Fig.~\ref{fig:fold}). Rather than instantiating nine copies, we implement one physical TCN engine and execute the blocks sequentially. What a spatial design hard-wires per block therefore becomes data that the engine fetches at each iteration: the output format of each operation, the block's weights, and its depthwise history.

\subsection{The shared data lane}

The folded engine must therefore preserve the different precisions learned across the network while reusing the same physical datapath. Activations therefore travel through the engine in a common fixed-point carrier format, and multiply--accumulate results are collected in a wider shared accumulator. At each operator boundary, the result is requantized to the output format learned during training. Rather than implementing a separate cast for every operator and activation element, the engine uses one shared requantizer controlled by a small cast ROM.

To supply this shared datapath, the weight lane stores consecutive weights for each output channel using a shared code width $b_f$ and shift $s_f$. For weight $j$ in channel $f$, the decoder reconstructs
\begin{equation}
    W_{f,j}=\operatorname{signext}(c_{f,j})\,2^{s_f},
\end{equation}
where $c_{f,j}$ is its $b_f$-bit signed code. The MAC therefore receives the same fixed-point weight as it would from the uniform-width representation; only its storage format changes. A channel with $b_f=0$ has only zero weights: it occupies no weight bits, and removing that channel from the schedule also removes its MAC cycles (Eq.~\ref{macestimate}).

Together, these representations preserve the learned fixed-point computation, but sharing also limits how much heterogeneous precision can reduce the arithmetic hardware. The shared multiplier must support the widest operand used anywhere in the engine, so a narrower weight shrinks the packed stream but not the multiplier. As a result, folding does not turn every per-weight bit-width reduction into a proportional reduction in compute hardware, as a fully spatial reading of effective bit operations (EBOPs) would suggest. The folded engine can still exploit two forms of compression directly: variable-width weight storage and exact-zero channels that are removed from the execution schedule.

\section{Training priced for the engine}\label{sec:trainprice}

All models minimize the compressed spectral MSE~\cite{braun2021} plus the two HGQ regularizers~\cite{hgq}. A monotone-$L_1$ term ($\lambda_b{=}10^{-8}$) pushes every learned bit width down uniformly, and $\beta\cdot$EBOPs sums $\mathrm{MACs}_\ell\, b^{(a)}_\ell b^{(w)}_\ell$ over layers (activation and weight widths), weighting precision by arithmetic cost. Our engine-aware recipes modify this baseline objective with terms for the two costs the folded hardware can actually exploit: weight storage and scheduled cycles.

For the folded engine, latency is set by the number of scheduled MACs; therefore, saving cycles requires removing computation. We encourage whole output channels to become zero so that their MAC loops can be skipped. For each body convolution $\ell$, let
$K^{(\ell)}_j$ denote all kernel weights producing output channel $j$.
We apply the cycle-weighted group-lasso
\begin{equation}
\mathcal{L}_{\mathrm{cycle}}
=
\frac{\lambda_C}{C_0}
\sum_{\ell \in \mathcal{B}}
\sum_j
c_\ell
\left\|K_j^{(\ell)}\right\|_2 ,
\label{eq:cycle}
\end{equation}
where $c_\ell$ is the number of MACs needed to compute one output channel of layer $\ell$, and $C_0$ normalizes the term by its initial value. Thus, channels that cost more scheduled cycles receive more pruning pressure. Setting $c_\ell = 1$ recovers the plain group-lasso~\cite{wen2016}.

Keeping this pressure active throughout training proved unstable. At epoch $50$, we therefore freeze the learned zero-channel mask and remove the fold-cost terms. A projection keeps masked channels at zero, so the hardware schedule can no longer change. The task loss, monotone-$L_1$, and $\beta\cdot$EBOPs terms remain active, allowing the surviving weights and precisions to continue adapting within the fixed sparse topology.

For the folded engine, reducing a weight width saves storage. We therefore learn one width $b_\ell$ per weight tensor and, when needed, directly constrain the packed weight tape. For a target capacity $B_0$, we add
\begin{equation}
\mathcal{L}_{\mathrm{tape}}
=
\lambda_T
\left[
\frac{B_{\mathrm{tape}}}{B_0}-1
\right]_+ ,
\label{eq:tape}
\end{equation}
where $B_{\mathrm{tape}}$ is the emitted weight-image size. The forward pass uses the exact packed size, including per-channel headers, while a differentiable estimate from the learned widths provides the gradient.

\section{Results}\label{sec:results}

\begin{table*}[t]
\centering
\caption{KV260 results for eight builds ($K{=}4$ MAC lanes). Tape size is in Mbit.}
\label{tab:mainresults}
\footnotesize
\setlength{\tabcolsep}{3.1pt}
\begin{tabular}{lllrrrrrrrrr}
\toprule
build & training objective & weight granularity & tape & zero MACs & LUT & BRAM & DSP & ms/frame & mW & $\mu$J/frame & PESQ \\
\midrule
Baseline F32 &  &  &  &  &  &  &  &  &  &  & 2.89 \\
\midrule
\textsc{Uniform-8}     & pinned width              & none        & $3.98$ & $0.1\%$  & $18.3$k & $141.5$ & $21$ & $1.85$ & $49.7$ & $92.1${\scriptsize$\,\pm\,$1.5} & $2.69$ \\
\textsc{Uniform-7}     & pinned width              & none        & $3.24$ & $0.2\%$  & $17.0$k & $121$   & $21$ & $1.83$ & $44.6$ & $81.5${\scriptsize$\,\pm\,$1.1} & $2.70$ \\
\textsc{Uniform-6}     & pinned width              & none        & $2.31$ & $0.7\%$  & $10.9$k & $114$   & $21$ & $1.77$ & $36.5$ & $64.5${\scriptsize$\,\pm\,$1.6} & $2.64$ \\
\midrule
\textsc{PerWeight-Wrap} & HGQ                                 & per-weight  & $3.22$ & $6.1\%$  & $19.4$k & $114.5$ & $21$ & $1.75$ & $42.8$ & $74.6${\scriptsize$\,\pm\,$1.4} & $2.74$ \\
\textsc{PerWeight-Sat} & HGQ                                  & per-weight  & $3.08$ & $6.2\%$  & $19.8$k & $110$   & $21$ & $1.74$ & $37.6$ & $65.4${\scriptsize$\,\pm\,$1.3} & $2.72$ \\
\midrule
\textsc{Store}         & HGQ $+$ storage                      & per-channel   & $2.98$ & $0.0\%$  & $20.7$k & $113$   & $20$ & $1.86$ & $44.5$ & $82.7${\scriptsize$\,\pm\,$2.4} & $2.76$ \\
\textsc{Skip}          & HGQ $+$ storage $+$ cycle lasso & per-channel & $1.97$ & $44.1\%$ & $21.7$k & $79$ & $21$ & $1.09$ & $34.1$ & $37.3${\scriptsize$\,\pm\,$1.1} & $2.69$ \\
\textsc{Skip-Plain}    & HGQ $+$ storage $+$ plain lasso & per-channel & $1.80$ & $47.3\%$ & $21.4$k & $79$ & $21$ & $1.03$ & $32.4$ & $\mathbf{33.4}${\scriptsize$\,\pm\,$1.6} & $\mathbf{2.78}$ \\
\midrule
\textsc{DEN16}$^\S$~\cite{olalere2026} & fixed $16$-bit, no compression & none & $22.5$ & --- & $90.9$k & $126.5$ & $899$ & --- & --- & --- & $2.41$ \\
\bottomrule
\end{tabular}\\[2pt]
\parbox{\textwidth}{\scriptsize All builds route at $100$\,MHz with $1.35$--$2.84$\,ns slack. Energy: mean\,$\pm$\,95\% CI over eight idle/active pairs. \textsc{DEN16}$^\S$~\cite{olalere2026} is an independently reported implementation of a different speech-enhancement network on the same KV260 FPGA.}

\end{table*}

\subsection{Experimental Setup}

\begin{description}
\item[Data]
We use the $16$\,kHz VoiceBank-DEMAND dataset~\cite{valentini}. Four of the $28$ training speakers (p279, p282, p286, p287) are held out as a speaker-disjoint validation set; the $824$-utterance test set (speakers p232, p257) is used only for final scoring.

\item[Models]
All builds use the same Conv-FSENet (Section~\ref{ssec:network}) and differ only in the training objective and the weight granularity they learn. The \textsc{Uniform} builds pin the total width of every quantizer, weights and activations alike, to $6$, $7$, or $8$ bits with $\beta{=}0$; only each quantizer's binary point is learned. \textsc{PerWeight-Wrap} and \textsc{PerWeight-Sat} use the stock HGQ objective at per-weight granularity, with wrapping and saturating data-lane overflow respectively. \textsc{Store} learns one width per tensor under the storage term (Eq.~\ref{eq:tape}, $B_0{=}3.0$\,Mbit). \textsc{Skip} learns per-channel widths and adds the cycle-weighted group-lasso (Eq.~\ref{eq:cycle}, $\lambda_C{=}0.02$) with the mask freeze at epoch $50$; \textsc{Skip-Plain} replaces cycle-weighted group-lasso with plain group-lasso $c_\ell{=}1$.

\item[Training]
The deployed checkpoint is the final epoch under an active compression schedule and, for mask-freeze runs, the validation-best checkpoint after projection; this rule was fixed before test scoring.

\item[Hardware]
All builds target the AMD Kria KV260 (Zynq UltraScale+ XCK26) through the same Vitis HLS/Vivado 2024.1 flow and run at $100$\,MHz. The folded engine of Section~\ref{sec:system} is driven from PYNQ~3.0.1 on the Cortex-A53 host, which also computes the STFT, the output sigmoid, and the inverse STFT.

\item[Bit-exactness]
An integer NumPy reference matches all $75$ layer outputs in the C simulation, including the cold start. On the board, replayed DMA inputs produce identical raw output words for every deployment; repeated runs over all $824$ test utterances are also identical. The host-side feature quantizer is checked separately against reference vectors, since DMA output parity does not verify it.

\item[Power and energy]
We measure KV260 board power (PS, PL, and DDR) from the SOM's $5$\,V INA260 sensor at $5$\,Hz. Each build is measured over eight interleaved pairs of $60$\,s idle and continuous-processing windows. Dynamic power is the mean active-minus-idle difference; energy per frame is this difference divided by the achieved frame rate. %
The idle board draws $3.11$--$3.13$\,W, about $50$\,mJ per real-time frame, so the energies reported are the accelerator's added share, under $0.2\%$ of the total.

\item[Metrics]
Quality is wideband PESQ (ITU-T P.862.2)~\cite{p8622} over all $824$ test utterances, computed with the mask network running on silicon. Frame time is board wall-clock per frame at the flat-out rate, and resource counts are post-route.
\end{description}

\subsection{Main comparison}

Table~\ref{tab:mainresults} compares eight models that all route and meet timing on the same engine. \textsc{Skip-Plain} gives the lowest measured latency and energy: $47.3\%$ of its convolution MACs belong to exact-zero channels, which the engine skips, yielding $1.03$\,ms and $33.4$\,\(\mu\)J per frame at PESQ $2.78$. That is $42\%$ lower latency and $48\%$ less consumption than the most efficient dense build, \textsc{Uniform-6}. Weighting the group-lasso by channel cycle cost does not improve this result. \textsc{Skip} skips $44.1\%$ of MACs and reaches $1.09$\,ms and $37.3$\,\(\mu\)J per frame at PESQ $2.69$, so the plain lasso performs better on all three measured outcomes in these runs.

The uniform and stock HGQ models have at most $6.2\%$ of MACs in skippable zero-width channels and run in $1.74$--$1.85$\,ms per frame. Pricing weight storage produces \textsc{Store}: its $2.98$\,Mbit tape ($4.46$ bits per weight) meets the $3.0$\,Mbit target and gives the highest quality among dense builds, but its $0.0\%$ skippable MACs leave the full compute schedule in place. Every build fits within the device's $144$ BRAM tiles, \textsc{Uniform-8} with $2.5$ to spare. %
It holds \textsc{Store} at $113$ tiles, $20\%$ fewer than \textsc{Uniform-8}, yet sparsity saves more: the sparse builds need $79$ tiles and finish far below $B_0$, where the hinge is inactive. %

Weight width also affects power: from eight to six bits in the pinned-width sweep, dynamic power falls by $27\%$ and energy per frame by $30\%$, while frame time changes by under $5\%$. Narrower weights thus save storage and, presumably through lower switching activity, dynamic energy, but not cycles.

DEN16~\cite{olalere2026} runs a different, time-domain network on the same board and dataset. Our builds score $0.23$--$0.37$ PESQ higher with $4.2$--$8.3\times$ fewer LUTs and $43\times$ fewer DSPs, the latter largely because nine TCN blocks share four MAC lanes. DEN16 is faster, however: its RTF is $0.010$ against our $0.064$, and its $9.7$\,ms first sample beats our $32$\,ms window. %

\begin{figure}[t]
\centering
\includegraphics[width=0.99\linewidth]{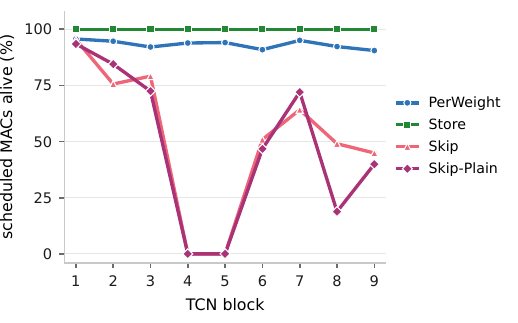}
\caption{Share of scheduled MACs in surviving output channels by TCN block.}
\label{fig:collect}
\end{figure}

\begin{table}[t]
\centering
\caption{MAC-lane sweep of \textsc{Skip-Plain}; outputs are byte-identical. $^\ast$Out-of-context synthesis.}
\label{tab:lanes}
\footnotesize
\setlength{\tabcolsep}{3pt}
\begin{tabular}{rrrrrrrrr}
\toprule
$K$ & ms/frame & speedup & lane eff. & mW & $\mu$J/frame & LUT & BRAM & DSP \\
\midrule
$1$  & $3.63$          & $1.00\times$ & $97\%$ & $26.1$ & $94.9$          & $20.9$k & $79.5$ & $8$ \\
$4$  & $1.03$ & $3.53\times$ & $86\%$ & $32.4$ & $33.4$          & $21.4$k & $79$   & $21$ \\
$8$  & $0.68$          & $5.34\times$ & $65\%$ & $43.8$ & $29.9$ & $25.8$k & $79$   & $37$ \\
$16$ & \multicolumn{3}{l}{does not meet $\mathrm{II}{=}1$}    & ---    & ---             & $43.2$k$^\ast$ & $116^\ast$ & $79^\ast$ \\
\bottomrule
\end{tabular}

\end{table}

Table~\ref{tab:lanes} varies \textsc{Skip-Plain} over $K \in \{1,4,8,16\}$ MAC lanes by changing only the unroll factors. Dynamic power rises from $26.1$ to $32.4$ to $43.8$\,mW, but energy per frame falls by $2.8\times$ and $1.12\times$ respectively. Lane efficiency falls for two reasons. First, some compute is done once per frame regardless of $K$: updating the depthwise state, controlling loops, and checking channels whose MACs will be skipped. Second, each lane decodes from a shared $(15K{+}64)$-bit window, so the decoder's barrel shifters grow as $K(15K{+}64)$ bit positions ($4\,864$ at $K{=}16$). At $K{=}16$, this logic prevents the convolution loop from starting a new iteration every cycle: its initiation interval rises to $\mathrm{II}{=}2$. %

\subsection{Effects of compression}\label{sec:collect}

To isolate the benefit of row skipping, we co-simulate \textsc{PerWeight-Wrap}, \textsc{Store}, \textsc{Skip}, and \textsc{Skip-Plain} with MAC skipping disabled. At $K{=}4$, the four builds take nearly the same number of cycles. Narrower weights therefore do not meaningfully shorten the full schedule on this engine.

With skipping enabled, \textsc{Skip-Plain} avoids $47.3\%$ of MACs and reduces frame time by $44.6\%$; \textsc{Skip} avoids $44.1\%$ of MACs and reduces frame time by $41.3\%$. The frame-time reductions are smaller because work outside the skippable MAC loops still runs, including frontend, mask, normalization, and residual operations (roughly $24$k cycles per frame).

Figure~\ref{fig:collect} shows that the training objectives produce different amounts of skippable computation. \textsc{Store} produces no such channels. Stock HGQ sets individual weights to zero throughout the network, but produces whole zero channels only in depthwise layers, accounting for $6.1\%$ of MACs. Channel group-lasso instead drives entire output channels to zero: both lasso builds remove TCN blocks $4$ and $5$ entirely and reduce the work in later blocks. In \textsc{Skip-Plain}, only $18.8\%$ of block $8$'s scheduled MACs remain.

\section{Conclusion}

In this paper, we show that compression chosen during training can reduce storage, latency, and energy on a folded FPGA engine for streaming speech enhancement. At the same four-lane configuration as the uniform-precision builds, \textsc{Skip-Plain} processes a frame in $1.03$\,ms and uses $33.4$\,$\mu$J, compared with $1.77$\,ms and $64.5$\,$\mu$J for \textsc{Uniform-6}. Its measured PESQ is $2.78$ versus $2.64$, so the hardware savings come at comparable or better measured quality.

Channel sparsity pressure during training drives entire output-channel weight rows to zero, allowing the engine to skip their MAC loops and reduce the work scheduled for each frame. Future work could explore alternative structured sparsity priors to reduce the training instability observed with Lasso regularization.

\clearpage

\bibliographystyle{IEEEtran}
\bibliography{refs_iscas}

\end{document}